\documentclass[showpacs,twocolumn,showkeys,amsmath,amssymb,prl,superscriptaddress]{revtex4}

\usepackage{amsmath,amssymb,amsfonts}
\usepackage{graphicx}
\usepackage{txfonts}
\usepackage{epstopdf}
\usepackage{mathtools}
\usepackage[T1]{fontenc}

\usepackage{framed}

\usepackage{color}

\usepackage[unicode]{hyperref}
\hypersetup{
   unicode=true,          
   a4paper=true,
   plainpages=false,
   pdftitle={Title of PDF},    
   pdfauthor={Author of PDF},     
   pdfsubject={Subject of PDF},   
   colorlinks=true,       
   citecolor=blue,        
}
\begin{document}

\title{Higher-order quantum
  bright solitons in Bose-Einstein condensates show truly quantum
  emergent behavior }

\author{Christoph Weiss}
\email{christoph.weiss@durham.ac.uk}
  \affiliation{Joint Quantum Centre (JQC) Durham--Newcastle, Department of Physics, Durham University, Durham DH1 3LE, United Kingdom}

\author{Lincoln D.\ Carr}
\email{lcarr@mines.edu}
\affiliation{Department of Physics, Colorado School of Mines, Golden, Colorado 80401, USA}

\date{\today}

 \begin{abstract}
When an interaction quench by a factor of four is applied to an attractive Bose-Einstein condensate, a higher-order quantum bright soliton exhibiting robust oscillations is predicted in the semiclassical limit by the Gross-Pitaevskii equation. Combining matrix-product state simulations of the Bose-Hubbard Hamiltonian with analytical treatment via the Lieb-Liniger model and the eigenstate thermalization hypothesis, we show these oscillations are absent.  Instead, one obtains a large stationary soliton core with a small thermal cloud, a smoking-gun signal for non-semiclassical behavior on macroscopic scales and therefore a fully quantum emergent phenomenon.
\end{abstract}
\pacs{	
05.60.Gg, 
03.75.Lm, 	
03.75.Gg 
}

\keywords{Bright soliton, Bose-Einstein condensation, Quantum
  many-body physics, Far-from-equilibrium quantum dynamics, Semiclassical breakdown}

\maketitle

The quantum-classical correspondence is well-established for single-particle quantum mechanics but is known to be problematic for some many-body quantum problems such as strongly correlated systems and even materials as simple as the antiferromagnet.  A key macroscopic prediction of Bose-Einstein condensates (BECs) is the bright soliton, appearing as a localized robust ground state ``lump'' for attractive BECs.  Based on the ubiquity of semiclassical limits for non-interacting and weakly interacting bosons, such as lasers and BECs, one might expect a well-defined emergent macroscopic classical behavior generically from such systems. To date, most aspects of matter-wave bright soliton experiments~\cite{KhaykovichEtAl2002,StreckerEtAl2002,CornishEtAl2006,EiermannEtAl2004,MarchantEtAl2013,MedleyEtAl2014,NguyenEtAl2014,EverittEtAl2015,MarchantEtAl2016,LepoutreEtAl2016,McDonaldEtAl2014}
seem to be explained on the semiclassical mean-field level via the Gross-Pitaevskii
equation (GPE): thus they display quantum behavior on a
single-particle level matching classical wave experiments such as nonlinear photonic crystals~\cite{sukhorukov1} and spin-waves in ferromagnetic films~\cite{kalinikos1998,kalinikos2000}.
This statement is supported by the fact that
\textit{quantum-quantum}\/ bright
solitons~\cite{CarterEtAl1987,LaiHaus1989,CastinHerzog2001,CarrBrand2004,CalabreseCaux2007b,SachaEtAl2009,MuthFleischhauer2010,BieniasEtAl2011}
--- matter-wave bright solitons that display quantum behavior beyond
the single-particle mean-field level ---
for many practical purposes show mean-field behavior predicted by the GPE emerging
already for particle numbers as low
as $N\gtrapprox 3$~\cite{MazetsKurizki2006}. So far, beyond-mean field effects only seem to
play a role if two or more distinct bright solitons are involved: two matter-wave \textit{quantum}\/
bright solitons can behave quite differently from matter-wave \textit{mean-field}\/ bright
solitons. Only the latter necessarily have a well-defined relative
phase~\cite{BillamWeiss2014}. Both the limit of well-defined phase~\cite{NguyenEtAl2014}
and the limit involving a  superposition of many phases~\cite{HoldawayEtAl2014,SakmannKasevich2016} are experimentally relevant for matter-wave
bright solitons~\cite{NguyenEtAl2014,SakmannKasevich2016}.   In this
Letter we show that truly quantum many-body effects are
responsible for the dynamics of a single quantum-quantum bright soliton, a smoking-gun signal for quantum emergence in BEC experiments.

For far-from
equilibrium dynamics of  beyond-ground state quantum bright solitons, we are only at
the beginning of a journey similar to the case of quantum
\textit{dark}\/ solitons.  That scientific voyage required multiple lines of
investigations~\cite{GirardeauWright2000,MishmashCarr2009,DelandeSacha2014,KronkeSchmelcher2014,KarpiukEtAl2015}
to arrive at the state-of-the art explanation that atom losses are
necessary to obtain  mean-field properties from many-body
quantum solutions~\cite{SyrwidSacha2015}.  Dark solitons were also
realized experimentally in BECs~\cite{BurgerEtAl1999} and have been further explored in detail over the years in comparison to such predictions, e.g.~\cite{weller2008}.  In contrast, bright solitons to-date lack for instance a phase coherence measurement, let alone the kind of far-from-equilibrium dynamics we are predicting here.  Thus we focus on a quantum bright soliton experiment easily accessible in current platforms.  Specifically, one first prepares a single ground-state bright soliton and then rapidly changes the interaction, an ``interaction quench'' via a Feshbach resonance, a well-established experimental technique.    For one-dimensional Bose gases recent work related to quenches includes
positive-to-negative
quenches~\cite{MuthFleischhauer2010,TschischikHaque2015},
and zero-to-positive quenches~\cite{ZillEtAl2014}.
Quenches involving dark-bright
solitons~\cite{SolnyshkovEtAl2016,LiuEtAl2016}, quenched dynamics of
two-dimensional solitary waves~\cite{ChenEtAl2013}, and
breathers in discrete nonlinear Schr\"odinger
equations~\cite{JohanssonAubry2000,KevrekidisEtAl2000,LahiriEtAl2000}
were also investigated, as well as the breathing motion after a quench of the strength of a harmonic
trap~\cite{BarbieroSalasnich2014}.

For attractive BECs, there are very specific mean-field predictions~\cite{CarrCastin2002}: in particular, for an interaction
quench by a factor of four there are exact analytical mean-field
results available that predict robust perfectly oscillatory behavior
for all times~\cite{SatsumaYajima1974}.
However, how \textit{quantum}\/ bright solitons would
behave in such a situation   is an open question which we address in
the current Letter.  One GPE interpretation of a higher order soliton is $N_s$ bound bright solitons, here $N_s=2$, a kind of diatomic solitonic molecule in a nonlinear vibrational mode.  One might therefore expect quantum fluctuations to cause the two solitons to unbind via e.g quantum tunneling out of a many-body potential, resulting in two equal-sized solitons moving away from each other~\cite{StreltsovEtAl2008}.  This is not at all what we find, and is inconsistent
with exact results for the center-of-mass wave
function~\cite{CosmeEtAl2016b}.  Moreover, our beyond-mean-field results are distinct from the GPE failing for strongly correlated systems like Mott
insulators~\cite{FisherEtAl1989,JakschEtAl1998,GreinerEtAl2002}; as well as from many-body systems on short timescales with differences disappearing for typical experimental parameters and large BECs~\cite{GertjerenkenWeiss2013}.  We will show that an interaction quench leaves a large soliton core with small emissions of single particles.  Experimentally these dynamics will appear as a ``fizzled'' higher order bright soliton, a stationary soliton core with a small thermal cloud.  Thus we establish a new kind of quantum macroscopicity in weakly interacting bosonic systems.

The mean-field approach via the GPE is a
powerful approximation which provides physical insight into weakly
interacting ultracold atoms. In a quasi-one-dimensional wave
guide~\cite{carr2000e,KhaykovichEtAl2002,StreckerEtAl2002,CornishEtAl2006,EiermannEtAl2004,MedleyEtAl2014,NguyenEtAl2014,EverittEtAl2015,MarchantEtAl2013,MarchantEtAl2016,LepoutreEtAl2016,McDonaldEtAl2014}
the GPE reads
\begin{equation*}
i\hbar \partial_t\varphi(x,t) = -(\hbar^2/2m)\partial_{xx}
\varphi(x,t)
+(N-1)g_{1 \rm D}|\varphi(x,t)|^2 \varphi(x,t),
\end{equation*}
 where  $\varphi(x,t)$ is a complex wave function normalized to unity and $N$ is the number of atoms of mass $m$. The  attractive interaction
\begin{align*}
g_{\rm 1D} &=2\hbar
\omega_{\perp}a <0
\end{align*}
 is proportional to the
\textit{s}-wave scattering
length $a$ and the perpendicular angular trapping-frequency,
$\omega_{\perp}$~\cite{Olshanii1998}.  Some GPE predictions for \textit{repulsive} BECs
 even become exact~\cite{LiebEtAl2000,ErdosEtAl2007} in the mean-field
 limit
\begin{equation}
\label{eq:MeanFieldLimit}
g_{\rm 1D}\to 0,\quad N\to \infty,\quad (N-1)g_{\rm 1D} = \rm const.
\end{equation}

While quantum bright solitons in their internal ground state in addition have a center-of-mass
wavefunction (see Refs.~\cite{GertjerenkenWeiss2012,DelandeEtAl2013} and
references therein), for measurements both many-body quantum
physics~\cite{CalogeroDegasperis1975,CastinHerzog2001}
 and
the GPE~\cite{PethickSmith2008} predict bright solitons
localized at $X_0$ with a
single-particle density profile of form
\begin{equation*}
\varrho(x)\equiv |\varphi(x)|^2 = (2\xi_N\left\{\cosh[(x-X_0)/(2\xi_N)]\right\}^2)^{-1},
\end{equation*}
where the soliton length $\xi_N$  and the related soliton time $\tau_N$
\begin{equation}
\label{eq:solitontimelength}
 \xi_N \equiv \hbar^2[m(N-1)\left|g_{\rm 1D}\right|]^{-1}; \quad \tau_N\equiv m\xi_N^2/\hbar.
\end{equation}
remain constant
 when approaching the mean-field limit~(\ref{eq:MeanFieldLimit}).

\begin{figure}
\includegraphics[width = \linewidth]{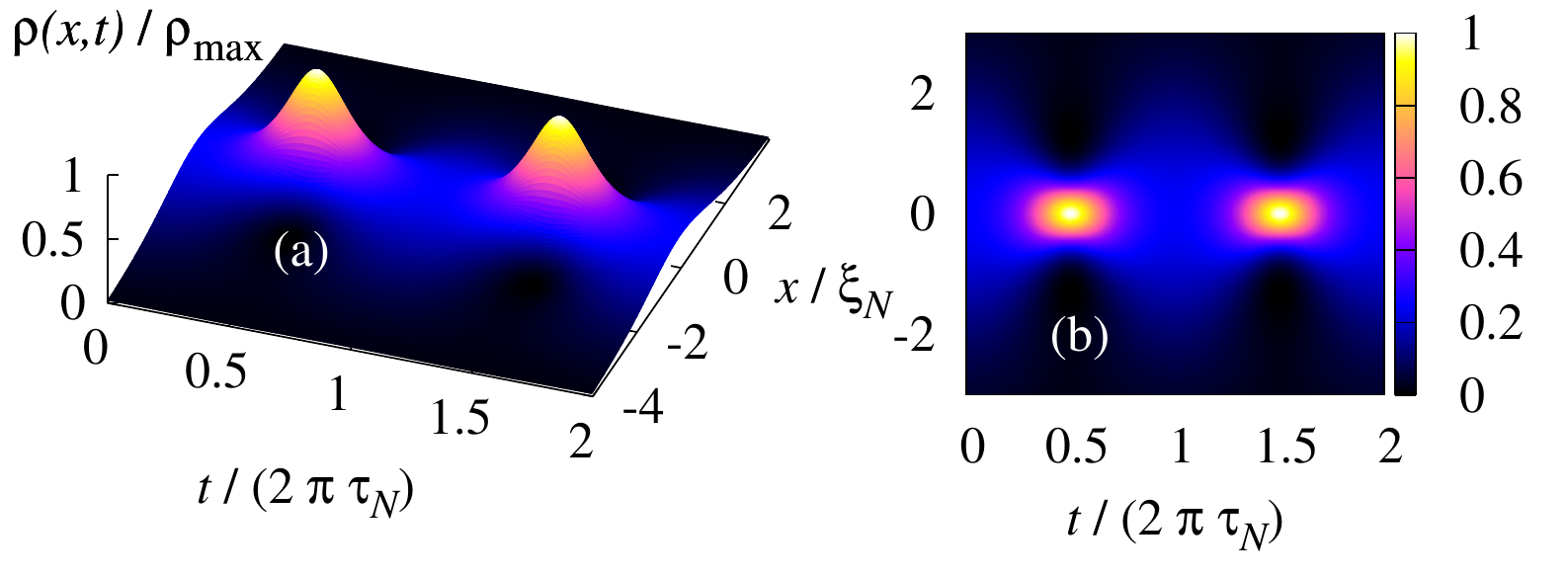}
\caption{\textit{Semiclassical emergent dynamics.} After an interaction quench by a factor of four, the GPE mean-field theory
  predicts perfect oscillatory behavior,
  Eq.~(\ref{eq:meanfieldrho})~\cite{SatsumaYajima1974}. Is it
  realistic to expect BEC experiments to reproduce these oscillations? (a) GPE density, normalized to its maximum for the
  first two oscillation periods as a function of space and time in soliton units. (b) 2D projection of (a).}
\label{fig:MeanField}
\end{figure}

In this Letter we use an interaction quench
\begin{equation*}
g_{\rm 1D} (t)= \left\{
\begin{array}{lcr}
g_0 &:&t\le 0\\
\eta g_0 &:&t>0
\end{array}
\right. ;\quad \eta\ge 1,\quad g_0<0.
\end{equation*}
After an interaction quench by a factor of $\eta=4$, the GPE yields the analytical
result~\cite[p~300]{SatsumaYajima1974}
\begin{equation}
\label{eq:meanfieldrho}
 \varrho(x,t) = \left|{\frac {\cosh \left[ 3x/(2\xi_N) \right] +3\,{{\rm e}^{-{\rm i}t/\tau_N}}\cosh \left[ x/
(2\xi_N) \right] }{3\,\cos \left( t/\tau_N \right) +4\,\cosh \left( x/\xi_N \right) +
\cosh \left( 2\,x/\xi_N \right) }}
\right|^2,
\end{equation}
which is depicted in Fig.~\ref{fig:MeanField}.
For $3/2< \eta^{1/2} < 5/2$ mean-field predictions also are
very specific: after losing a few atoms the system self-cools to the robust higher-order
bright soliton of Eq.~(\ref{eq:meanfieldrho})~\cite{CarrCastin2002}.

Within the text book derivation~\cite{PethickSmith2008} of the GPE, the
many-body wave function corresponding to the GPE is a
Hartree-product state
\begin{equation}
\label{eq:Hartree}
\psi_{\rm GPE}(x_1,x_2,\ldots,x_N) = \textstyle\prod_{j=1}^N\varphi(x_j).
\end{equation}
The mean-square difference between the position of two
particles
\begin{equation*}
\langle(x_1-x_2)^2\rangle_{\rm GPE}(t)\equiv\textstyle\int\! dx_1\! \int dx_2\,
\varrho(x_1,t) \varrho(x_2,t)(x_1-x_2)^2,
\end{equation*}
a measure that is distinct from and independent of the
expansion of the center-of-mass wave function, can
thus be calculated form the  above analytical result to obtain
\begin{equation}
\label{eq:Delta12}
\Delta_{1,2} \equiv \langle(x_1-x_2)^2\rangle(t)/\langle(x_1-x_2)^2\rangle(0).
\end{equation}
While the mean-field prediction thus is a perfectly periodic function
of period $2\pi\tau_N$, the question is what we expect to find
on the
many-body quantum level  .

For fundamental considerations on the many-body level
corresponding to the GPE, a very
useful tool is  the Lieb-Liniger Model (LLM) with the Hamiltonian~\cite{LiebLiniger1963,McGuire1964} \begin{equation}
\label{eq:LL}
\hat{H} = -(\hbar^2/2m)\textstyle\sum_{j=1}^N(\partial^2/\partial x_j^2)+\sum_{j=1}^{N-1}\sum_{\nu=j+1}^{N}g_{\rm 1D}\delta(x_j-x_{\nu}),
\end{equation}
where $x_j$ denotes the position of particle $j$ of mass $m$. The ground-state energy is given by \begin{equation*}
E_0(N)= -mg_{1\rm D}^2 N(N^2-1)/(24\hbar^2).
\end{equation*}
The energy eigenstates of excited states can be written as
\begin{equation}
\label{eq:Eex}
E=\textstyle\sum_{r=1}^{N_{\rm S}}\left(E_0(N_r) +N_r \hbar^2 k_r^2/2m\right), \:\:
N=\sum_{r=1}^{N_{\rm S}} N_r,\:\: N_r>0
\end{equation}
corresponding to the intuitive interpretation of $N_{\rm S}$ solitonlets --- solitons
that contain a fraction of the total number of particles --- of
size~$N_r$ ($r=1,2,\ldots,N_{\rm S}$) and their individual center-of-mass kinetic
energy. Equation~(\ref{eq:Eex}) is valid if the system size $L$ is
large compared to even a two-particle soliton --- this can be included
by adding a diverging system size to  the mean-field
limit~(\ref{eq:MeanFieldLimit}) to get~\cite{Weiss2016,HerzogEtAl2014}
\begin{equation*}
g_{\rm 1D}\to 0,\; N\to \infty,\; L\to\infty,\; \xi_N = {\rm const.},\; N/L= {\rm const.}
\end{equation*}
Reaching
such a limit is a difficult numerical
problem~\cite{AlcalaEtAl2016}. However, by replacing the
Hamiltonian~(\ref{eq:LL}) by the Bose-Hubbard model (BHM) used to
model quantum bright solitons by e.g.~\cite{GertjerenkenWeiss2012,BarbieroEtAl2014,GertjerenkenKevrekidis2015,DelandeEtAl2013}, we introduce thermalization mechanisms present in real
experiments such as a weak imperfectly harmonic trap, or a
one-dimensional waveguide embedded in a 3D
geometry; for the BHM thermalization is due specifically to a lattice, in our case in the limit of very weak discretization.  The
BHM takes the form
\begin{align}
\label{eq:BH}
\hat{H}_{\rm BHM} =~&-J\textstyle\sum_{j}\left(\hat{b}_j^{\dag}\hat{b}_{j+1}^{\phantom{\dag}}
+\hat{b}_{j+1}^{{\dag}}\hat{b}_{j}^{\phantom{\dag}}\right)
+  \frac{1}{2}U\sum_{j}\hat{n}_j\left(\hat{n}_j-1\right)
\end{align}
where $\hat{b}_j^{\dag}$ ($\hat{b}_{j}^{\phantom{\dag}}$) creates
(annihilates) a particle on lattice site $j$, $U<0$ quantifies the
interaction energy of a pair of atoms and $\hat{n}_j$ counts the
number of atoms on lattice site $j$. In order to use this in a way we
can directly use the physical insight gained from the LLM~(\ref{eq:LL}), we choose for the hopping matrix element   (cf.~\cite[Eq.~(17)]{GertjerenkenWeiss2012})
\begin{equation*}
J = \hbar^2/(2m {\delta_{\rm L}}^2)
\end{equation*}
such that both models have
the same single-particle dispersion in the long wavelength limit
$k\delta_{\rm L}\ll \pi$, with $\delta_{\rm L}$ the lattice constant. The interaction
\begin{equation*}
U = g_\mathrm{1D}(32\,J{{\hbar}}^{2}m+{m}^{2}{g}^{2}_\mathrm{1D})^{1/2}/(4{{\hbar}}^{2})
\end{equation*}
is chosen such that the two-particle ground state has the same
 ground-state energy
 as Eq.~(\ref{eq:LL}) compared to the free gas~\cite{GertjerenkenWeiss2012}.

While both the weak lattice
introduced by Eq.~(\ref{eq:BH})] and a weak harmonic trap~\cite{MazetsSchmiedmayer2010,GringEtAl2012} break the
integrability of the LLM, we can still approximately describe
eigenstates by Eq.~(\ref{eq:Eex}).
Furthermore, from a modeling point of view, by choosing the lattice we avoid the divergence of the energy
fluctuations of the initial state immediately after the interaction
quench, caused by delta functions squared, in $\langle \Delta \hat{H}^2_{\rm new}\rangle_0 = (\eta -1)^2\left[
 \langle \hat{H}_{\rm  int~old}^2\rangle_0-\langle\hat{H}_{\rm
   int~old}\rangle_0^2\right]$. While in physics distributions with
well-defined mean and diverging variance are
well-known~\cite{ZaburdaevEtAl2015}, a more severe reason for avoiding
the LLM limit is that this limit seems to be mathematically
ill-defined -- an
initial wave function with the wrong boundary conditions at
$x_j=x_{\ell}$ ($j\ne\ell$) has to be expressed in terms of eigenfunctions with the
correct boundary conditions~\cite{CastinHerzog2001}.
Summarizing, we note that these energy fluctuations are consistent with the LLM
predicting the presence of quantum superpositions involving many
solitonlets in the initial state,   but inconsistent with simple pictures
predicting two large solitonlets that either oscillate~\cite{SatsumaYajima1974} around each
other or separate from each other~\cite{StreltsovEtAl2008} as the
latter cannot happen rapidly~\cite{CosmeEtAl2016b}.

We use time-evolving block decimation
(TEBD)~\cite{Vidal2004} --- a numerical method based on matrix product states~\cite{Schollwock2005,WhiteFeiguin2004}  --- to solve
the BHM~(\ref{eq:BH}).
In order to
exclude both boundary effects and effects introduced by additional
traps, we start with a very weak harmonic trap, such that opening it
hardly introduces atom losses~\cite{Castin2009} and
thus the analytical result~(\ref{eq:meanfieldrho}) remains valid.
In our simulations, we switch this trap off at the same time as we introduce the interaction
quench.

\begin{figure}
\includegraphics[width=\linewidth]{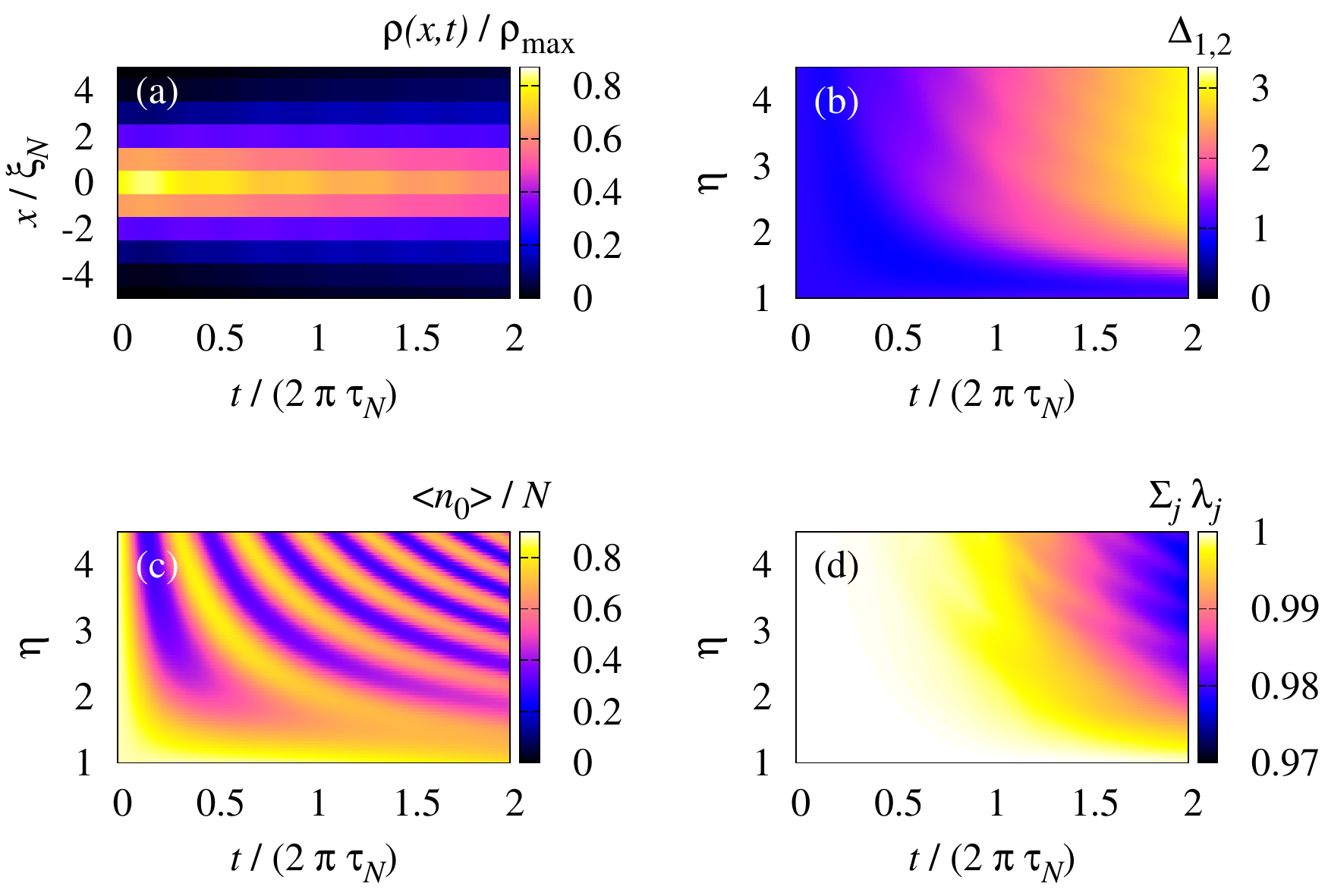}
\caption{\textit{Many-body quantum emergent dynamics. Following an interaction
  quench by a factor of $\eta$, the soliton} shows no indication of special values
  for the strength of the quench $\eta$ (initial parameters used for BHM:
  $N=3$, $J=0.5$, $U\simeq-0.5$). (a) Single
  particle density as a function of both position and time;
the mean-field oscillations displayed in Fig.~\ref{fig:N3} for $\eta
  =4$ are absent in the TEBD data. (b) Square root of relative width $\sqrt{\Delta_{1,2}}$   as a function of both time and
  quench strength $\eta$ shows that the absence of mean-field
  oscillations are generic. (c) The condensate fraction provides a
  further indicator of beyond-mean field behavior. (d) The sum over
  the largest 11 eigenvalues $\lambda_j$ of the single-particle density matrix
  (normalized to 1) shows that the system becomes even less mean-field
  for longer times.}
\label{fig:N3}
\end{figure}

\begin{figure}
\includegraphics[width=\linewidth]{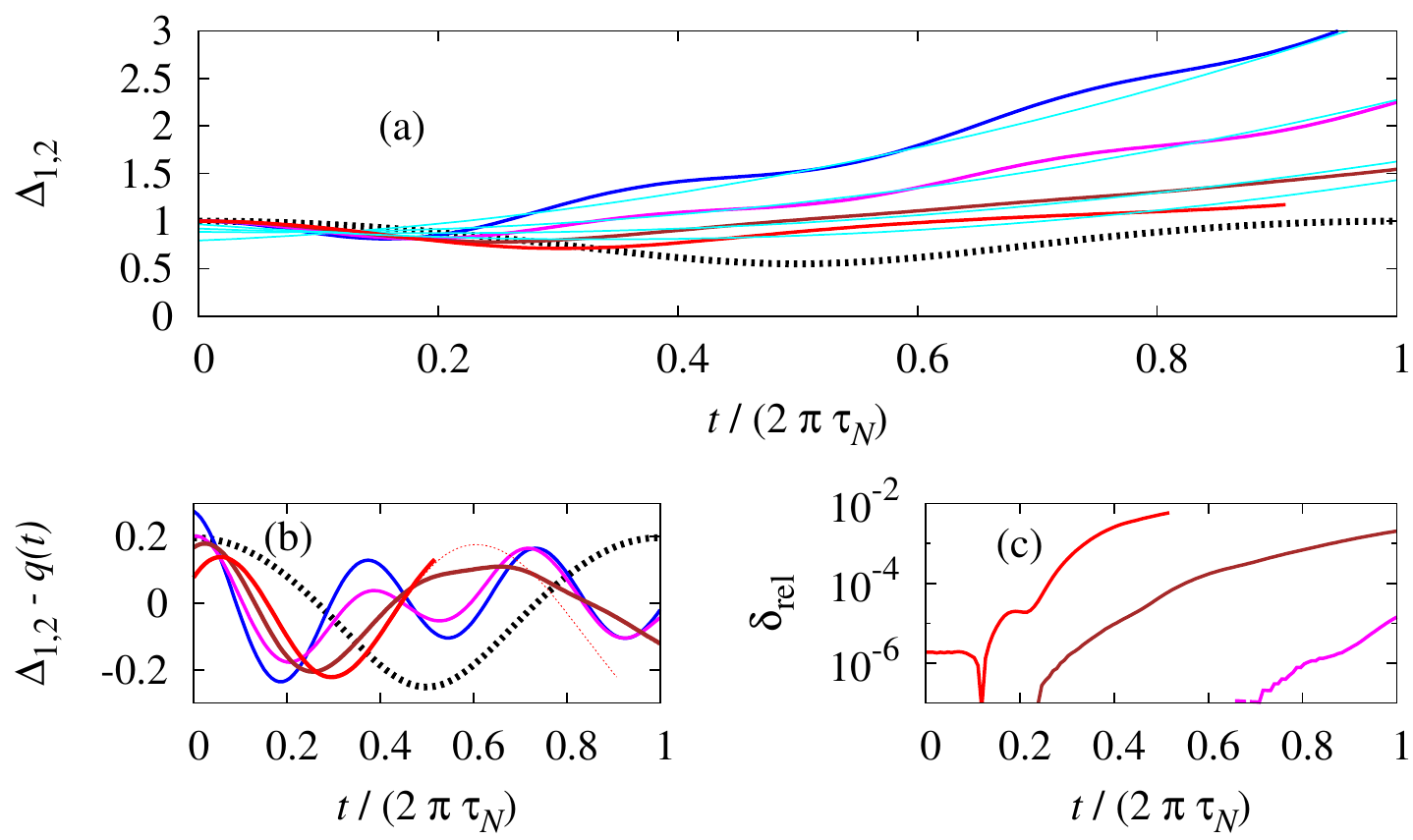}
\caption{\textit{Relative width measures for up to 16 particles.} While the GPE predicts perfect oscillations, the predominant behavior in the matrix-product state numerics is that the relative distance of particles grows.
(a) Relative width ${\Delta_{1,2}}$
  as a
function of time: mean-field oscillations (lowest black dotted curve); TEBD results ($N=3,\,4,\,8,\,16$ blue, fuscia, brown, red curves from top to bottom); overall behavior is quadratic (fits shown as light blue curves).  Here $J=0.5$ and $U\simeq \,-0.5,\,-0.33,\,-0.14,\,-0.07$. (b) Residual oscillations in panel (a) after subtracting the quadratic fit.  (c) Relative error
comparing TEBD data with distinct convergence parameters. From top to
bottom: $N=16$: $\chi_{max}= 60$ versus $\chi_{max}= 80$, $N=8$:
$\chi_{max}= 60$ vs.\ $\chi_{max}= 80$, $N=4$: $\chi_{max}= 30$ vs.\
$\chi_{max}= 40$, $N=3$ (too small to be visible):  $\chi_{max}= 30$ vs.\
$\chi_{max}= 40$.
}
\label{fig:largeN}
\end{figure}

If for $N\gtrapprox 3$ quantum bright solitons indeed already show
mean-field behavior~\cite{MazetsKurizki2006}, we should be able to see
the mean-field oscillations depicted in Fig.~\ref{fig:MeanField}
already for $N=3$.
Figures~\ref{fig:N3} and \ref{fig:largeN} show that
the mean-field oscillations are absent from the many-body TEBD data for $N=3$ and $N$ up to 16, respectively.  For the BHM~(\ref{eq:BH}), we note that the relative particle measurement of Eq.~(\ref{eq:Delta12}) can be rewritten in second quantized form appropriate to TEBD by
replacing \mbox{$\langle(x_1-x_2)^2\rangle$} with
$\sum_{j,\ell}(j-\ell)^2\langle
\hat{b}_j^\dag\hat{b}_j^{\phantom{\dag}}\hat{b}_\ell^\dag\hat{b}_\ell^{\phantom{\dag}}\rangle$.


Within
 the LLM, for relative distances large compared to the
 soliton length $\xi_N$, the
 leading-order contributions to excited states consists of
$N_{\rm   S}$ solitonlets of terms
 corresponding to $N_{\rm   S}$ individual solitonlets moving apart~\cite{CastinHerzog2001}
  In order to obtain a physical understanding on the time scales on
which these solitonlets can move apart if they initially sit on top
of each other, we recall the text book result for the variance of
an initially Gaussian
single-particle wave function~\cite{Fluegge1990},
\begin{equation*}
\Delta X^2 = \Delta X_0^2\{1+[\hbar t/(2M\Delta X_0^2)]^2\}.
\end{equation*}
For the relative motion the mass $M=Nm$ has to be replaced by the relative mass $m_{\rm rel}\equiv m\,N_1N_2/(N_1+N_2)$.
 If initially localized to $\Delta X_0\propto \xi_N$ (much stronger
 localization leads to too high kinetic energies while much weaker
 localization leads to a too wide initial wave function) and for a
 relative mass independent of $N$, the relative wave function will
 expand to a size larger than the initial wave function on time scales
 [cf.~Eq.~(\ref{eq:solitontimelength})]
\begin{equation}
\label{eq:ImportantTime}
t\propto m_r\xi_N^2/\hbar = [(N_1N_2)/(N_1+N_2)]\tau_{N}.
\end{equation}

The Hartree product states~(\ref{eq:Hartree}) are also ideal to calculate mean energies
which in the mean-field limit~(\ref{eq:MeanFieldLimit}) become
identical to the exact many-body quantum
results~\cite{CastinHerzog2001}. The kinetic energy prior to the
interaction quench is $\langle E_{\rm kin}\rangle_{\rm old}=N^3mg_{1\rm D}^2/(24\hbar^2) = -
E_0^{\rm old}$, and the interaction energy
$\langle E_{\rm int}\rangle_{\rm old}= 2E_0^{\rm old}$. Immediately after the interaction quench, the kinetic
energy remains unchanged and the interaction energy is increased to
$\langle E_{\rm int}\rangle_{\rm new}=\eta \langle E_{\rm
  int}\rangle_{\rm old}=2\eta E_0^{\rm old}$.
 In units of  the new ground state energy
$E_0^{\rm new} = \eta^2 E_0^{\rm old}$
we have a total average energy after the interaction quench of
\begin{equation}
\label{eq:Eaverage}
{\langle E\rangle}= [(2\eta-1)/\eta^2] {E_0^{\rm new} }
\end{equation}
where $0< (2\eta-1)/\eta^2<1$ for $\eta>0.5$ and $\eta\ne 1$.

If ultracold attractive atoms are initially prepared in their ground
  state, by using the eigenstate thermalisation
  hypothesis~\cite{PolkovnikovEtAl2011,EisertEtAl2015}  we conjecture that an interaction quench by a factor of
  $\eta$ will on short time scales lead to a final state consisting
  of a single bright soliton containing $N_1$ atoms, as given by
  thermodynamic predictions, and $N-N_1$ free atoms. In the mean-field
  limit~(\ref{eq:MeanFieldLimit}), for bright solitons
  thermodynamic predictions read~\cite{WeissGardinerGertjerenken2016}
\begin{equation}
\label{eq:target}
N_1 =\left(\frac{2\eta-1}{\eta^2}\right)^{1/3}N,\;\;  \xi_{N_1} =
\frac1{[\eta(2\eta-1)]^{1/3}}\xi_N, \;\; \eta>\frac 12,
\end{equation}
i.e., one large
soliton with reduced particle number $N_1$ and reduced size
$\xi_{N_1}$; and $N-N_1$ single atoms which are not bound in
molecules.

To suggest that this might indeed be what happens seems
counterintuitive at best, since following ``thermalization'' according to
the eigenstate thermalization hypothesis, all energetically accessible
eigenfunctions will be involved~\cite{PolkovnikovEtAl2011,EisertEtAl2015}; violating both the Landau
hypothesis~\cite[very end]{LandauLifshitz2002b}, which at first glance
seems to prevent co-existence of a large soliton and a free gas;
argued against also by mean-field predictions~\cite{SatsumaYajima1974,CarrCastin2002} as
well as thermodynamic predictions for ultracold atoms in contact with
a heat bath~\cite{HerzogEtAl2014,Weiss2016}. However,
the Landau hypothesis  is based on
assumptions that are not fulfilled for bright
solitons~\cite{WeissGardinerGertjerenken2016} and thermally isolated
ultracold atoms,  arguably realised in state-of-the-art experiments
with bright solitons~\cite{KhaykovichEtAl2002,StreckerEtAl2002,CornishEtAl2006,EiermannEtAl2004,MarchantEtAl2013,MedleyEtAl2014,NguyenEtAl2014,EverittEtAl2015,MarchantEtAl2016,LepoutreEtAl2016,McDonaldEtAl2014}, behave quite differently from those in contact with a
heat bath~\cite{WeissGardinerGertjerenken2016}.
Furthermore, contrary to rumours stating otherwise, one-dimensional
Bose gases do thermalise,  for example in the presence of a weak
harmonic trap~\cite{MazetsSchmiedmayer2010,GringEtAl2012}.

\begin{figure}
\includegraphics[width = \linewidth]{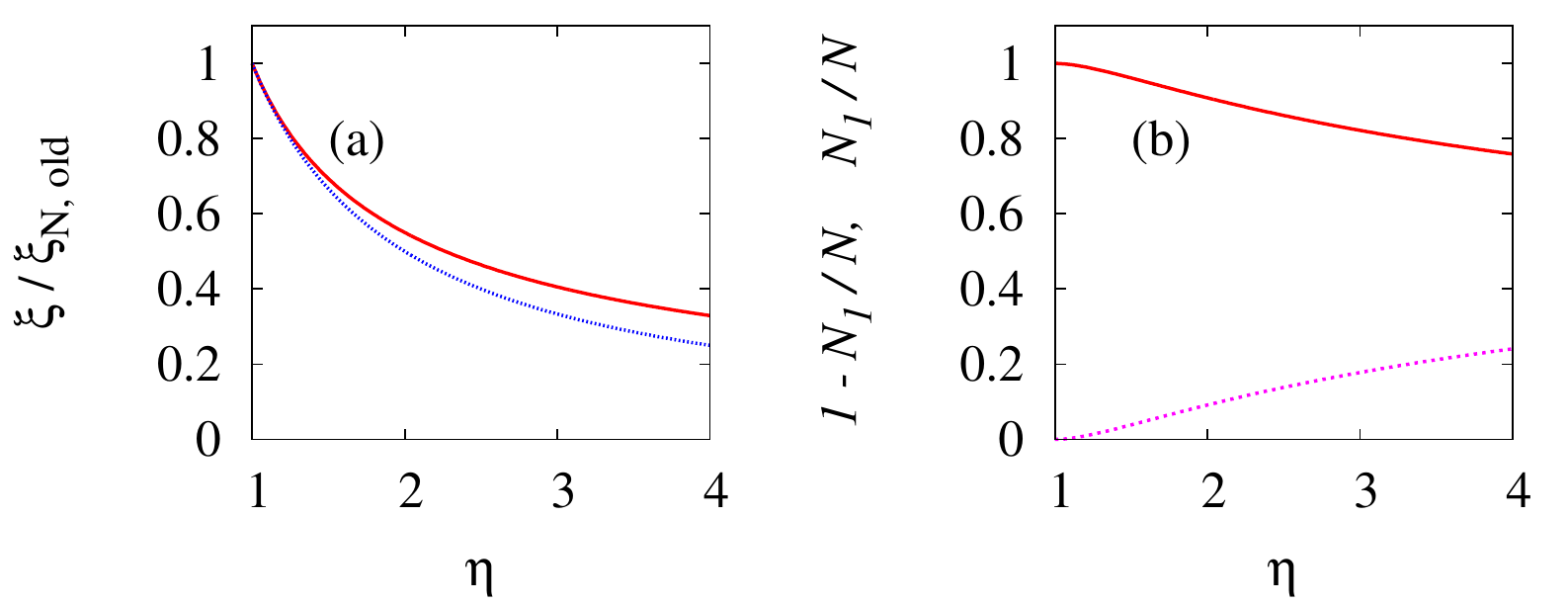}
\caption{\textit{Extrapolation to large particle number.} By applying the eigenstate thermalisation
  hypothesis~\cite{PolkovnikovEtAl2011,EisertEtAl2015} we conjecture
  that after an interaction quench to more negative interactions by a
  factor of $\eta$, the attractive BEC approaches the equilibrium
  predictions for a thermally isolated gas of Ref.~\cite{WeissGardinerGertjerenken2016}.
(a) New soliton length, in units of the original soliton length $\xi_{N,\rm old}$,
for a soliton containing all atoms (lower curve) versus $N_1$ emitted free atoms
(upper curve) as a function of the quench~$\eta$. (b) Fraction of atoms in the soliton (upper
curve) and in the free gas (lower curve) [see also Eq.~(\ref{eq:target})].
}
\label{fig:conjecture}
\end{figure}

As depicted in Fig.~\ref{fig:conjecture}, we conjecture
that after an interaction quench to more negative interactions, the
system will relax towards the situation predicted in thermal
equilibrium: the co-existence between one large soliton and a free
gas~\cite{WeissGardinerGertjerenken2016}. The application of the eigenstate thermalisation
  hypothesis~\cite{PolkovnikovEtAl2011,EisertEtAl2015} is supported by
  the fact that single atoms will move the initial cloud much faster
  than larger solitonlets [Eq.~(\ref{eq:ImportantTime})] that can
  continue to thermalize. Two macroscopic solitons sitting on top of
  each other would have to
  remain at the same position~\cite{CosmeEtAl2016b}; a freely
  expending gas passes the convergence test of
  Ref.~\cite{CosmeEtAl2016b} but energy conservation
and Eq.~(\ref{eq:Eaverage})
 would require at
  least one soliton(let) to be present.

To conclude, we have combined evidence from three distinct models
(GPE, LLM and BHM) to show that truly quantum emergent behaviour for
attractive Bosons happens after an interaction quench to more attractive
interactions.
Combining the numerical evidence with general
considerations based on the eigenstate thermalization hypothesis for
larger particle numbers, we conjecture that the final many-body
quantum state consists of one smaller bright soliton and lots of
single atoms, thus yielding
an ultimate example of
a mean-field breakdown on time scales that remain experimentally relevant even in the
mean-field limit~(\ref{eq:MeanFieldLimit}).   Our predictions are accessible to state-of-the-art experiments
with thousands of
atoms~\cite{KhaykovichEtAl2002,StreckerEtAl2002,CornishEtAl2006,EiermannEtAl2004,MedleyEtAl2014,NguyenEtAl2014,EverittEtAl2015,MarchantEtAl2013,MarchantEtAl2016,LepoutreEtAl2016,McDonaldEtAl2014}.
  Furthermore, the above
  conjecture offers an explanation as to why experiments that quasi-instantaneously
  switch from repulsive to attractive interactions (see for example
  Ref.~\cite{MarchantEtAl2013}) while avoiding the ``Bose-nova'' collapse or modulational instability can nevertheless lead to
  one large matter-wave bright soliton (and a thermal cloud) being formed.

\acknowledgments
We thank T.~P.~Billam, J.~Brand, J.~Cosme, S.~A.~Gardiner, B.~Gertjerenken,
and M.~L.~Wall for discussions. We thank the
Engineering and Physical Sciences Research Council UK for funding (Grant No.~EP/L010844/).  This  material  is  based in  part  upon  work  supported  by  the  US National  Science
Foundation  under  grant  numbers  PHY-1306638,  PHY-1207881, and PHY-1520915, and the US Air Force Office of Scientific Research grant number FA9550-14-1-0287.  L.D.C. thanks Durham University and C.W. the Colorado School of Mines for hosting visits to support this research.

Data will be available online soon~\cite{WeissCarr2016Data}.

\end{document}